\documentclass[conference]{IEEEtran}
\usepackage[utf8]{inputenc}
\usepackage{amsmath}
\usepackage{amsfonts}
\usepackage{amssymb}
\usepackage{graphicx}
\usepackage[left=2cm,right=2cm,top=2cm,bottom=2cm]{geometry}
\usepackage{blindtext, graphicx}
\usepackage{url}

\usepackage{xcolor}
\usepackage{colortbl}
\usepackage{lipsum}

\usepackage{colortbl,xcolor}

\usepackage{caption}

\ifCLASSINFOpdf
\else
\fi

\begin{document}
%
\title{Smart Car Privacy: Survey of Attacks and Privacy Issues}

\author{\IEEEauthorblockN{Akshay Madhav Deshmukh}
\IEEEauthorblockA{Department of Informatik\\
Technische Universit\"at Darmstadt\\
64289, Darmstadt\\
Email: madhavdeshmukh.akshay@stud.tu-darmstadt.de}}

\maketitle

\begin{abstract}
\par Automobiles are becoming increasingly important in our day to day life. Modern automobiles are highly computerized and hence potentially vulnerable to attack. Providing many wireless connectivity for vehicles enables a bridge between vehicles and their external environments. Such a connected vehicle solution is expected to be the next frontier for automotive revolution and the key to the evolution to next generation intelligent transportation systems. Vehicular Ad hoc Networks (VANETs) are emerging mobile ad hoc network technologies incorporating mobile routing protocols for inter-vehicle data communications to support intelligent transportation systems. Thus security and privacy are the major concerns in VANETs due to the mobility of the vehicles. Thus designing security mechanisms to remove adversaries from the network remarkably important in VANETs.
\par This paper provides an overview of various vehicular network architectures. The evolution of security in modern vehicles. Various  security and privacy attacks in VANETs with their defending mechanisms with examples and classify these mechanisms. It also provides an overview of various privacy implication that a vehicular network possess.
\end{abstract}


\section{Introduction}
\par Vehicles have advanced into complex systems incorporating many different Electronic Control Units (ECUs) which are interconnected based on the required functionality of different tasks. ECUs and a set of sensors are used to gather information about the vehicle’s behaviour and environment, and to control the functionalities of the vehicle. These ECUs are associated by exchanging messages resulting in an in-vehicle network also known as on-board network.  Thus, internal network infrastructure in vehicles has advanced in providing different use cases for different properties through a common bus system. 

\begin{figure}[h!]
\centering
\includegraphics[scale=0.25]{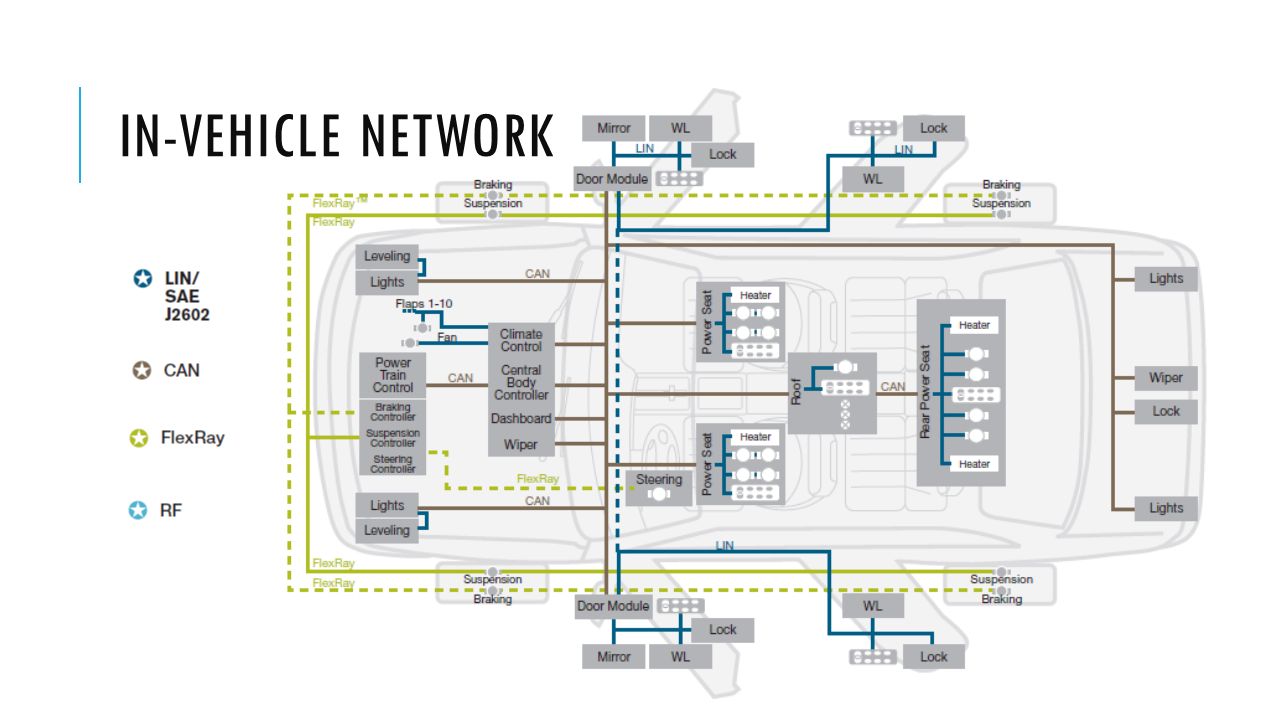}
\caption{In-Vehicle Network \cite{eeworld2021protocols}}
\label{fig:In-Vehicle Network}
\end{figure}

\par Vehicles in motion communicate with their neighbouring vehicles (closer vicinity vehicles) and with Road Side Units (RSUs) either directly or indirectly - through intermediate nodes. Each vehicle updates its neighbouring RSUs with its own information (speed, location etc.) and traffic updates on a regular bases. These RSUs and vehicles constitute Vehicular Ad-hoc Networks (VANets) also known as inter-vehicle Network to have the shared knowledge of the on going traffic updates\cite{4977227}

\par Furthermore, today's vehicles provide connectivity bridge to the outside world through wired and wireless interfaces. A Wireless Personal Area Network (WPAN) devices that comprise of wireless technologies which communicate through short range (upto 100m) such as Bluetooth, USB, Near Field Communication (NFC), Cellular networking, onboard diagnostic ports (ODBs) and many more. WPAN gateway provides a medium for communication between personal devices and in-vehicle ECUs. For example the driver can control the lights, windshield wipers, air flow, heat control, automatic ignition of the engine, automatic door lock systems and various other features through a bluetooth connected PDAs\cite{1634936}.

\par Vehicles that are connected to mobile devices can communicate with Service Centers (SCs) through mobile network (cellular network). Service Centers exchange information with RSUs by providing the required information about their location, behaviour and environment to the vehicle owners. The cellular network provides communications to the devices that have wireless communication capabilities with mobile and landlines\cite{zhang2005handbook}.

\par These technologies have their own advantages in providing useful services and functionalities, but the main concern seems to be the degree of vulnerability that a system involves.  It provides the room for intruders to carry out a potential attack on the system. In particular, with wireless interfaces to the outside world open up possibilities for remote attacks. 

\par Considering the new attack surfaces, it is very important to reconcile that security mechanisms are a vital aspect to add to the on-Board (in-vehicle networks) infrastructure. There are many state of art research proposals on reinforcing the automotive security based on cryptographic mechanisms that provide manipulation and authentication to solve bus security issues\cite{wolf2004security} and establishing an authentic confidential communication between the components of the vehicle and authentic controllers (authentic certified users)\cite{5164434}. Nevertheless, the biggest challenge of automotive security is its long life cycle of vehicles and by their safety requirements which mandates a traditional approach to make changes in the deployed vehicles. Thus, every changes introduced in the safety-critical system should be regression free and flawless.

\subsection{Security in modern vehicles}
\par Security in modern vehicles can be classified into on-Board (in-vehicle) security and V2X (Vehicle-to-Everything) communication security. V2X security infrastructure in turn are classified into Vehicle-to-vehicle communication security  and Vehicle-to-infrastructure security. 

\begin{figure}[h!]
\centering
\includegraphics[scale=0.28]{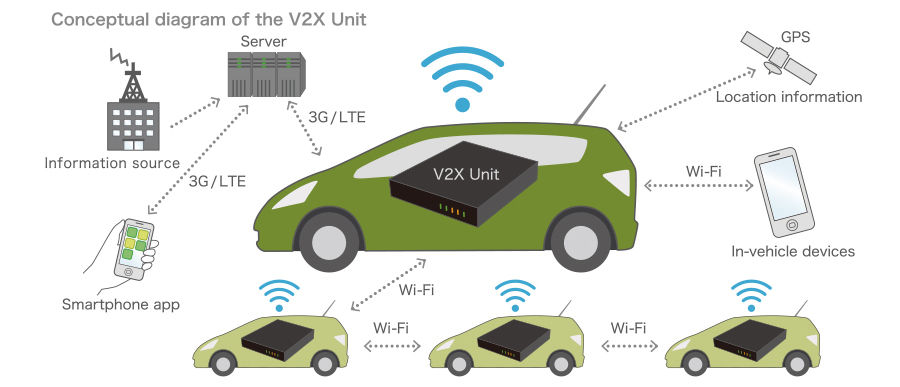}
\caption{V2X Architecture - Conceptual Level \newline\footnotesize source: http://world.honda.com/safety/hearts/2016/14/}
\label{fig:V2X Architecture - Conceptual Level}
\end{figure}

\par All the modern vehicles are embedded with ECUs which are categorised based on their functionalities-Powertrain Control Module (PCM), Transmission Control Module (TCM), Brake Control Module (BCM), Central Control Module (CCM), Central Timing Module (CTM), General Electronic Module (GEM), Body Control Module (BCM), Suspension Control Module (SCM). With respect to the comfort of the owner, a vehicle can comprise of a system for infotainment (e.g. radio and navigation system) and telematics units through a cellular network. Further, V2X capabilities can be broadly enhanced with vehicular adhoc networks (VANETs)\cite{othmane2015survey}.
\par Conversely, the on-board system acts as the attack surface for controlling the core aspects of the vehicle which involves safety-critical functionalities such as brake system and throttle (fueling system) by withdrawing arbitrary data by reading their memory from ECUs and feeding ECUs with malicious code that creates a havoc\cite{miller2015remote}. 
\par As VANETs are young, and still  under test beds, there are no real attacks documented as such. The IEEE standard specifies to embed vehicles with certificates with signature messages (Public Key Infrastructure) among the communicating partners to guarantee integrity within the system\cite{6509896}.
\par On-board  vehicular systems should be provided with integrity, authentication, authorization, and availability in order to safeguard against malicious attacks on ECU software and unauthorized spoofing of messages.Thus, research has initiated to secure the Controller Area Network (CAN) itself and provide a secure on-board system architecture\cite{kleberger2014towards},\cite{Wolf2007}. 

\par In V2I scenarios, when the vehicle wants to get the access to services like internet service or information about the nearest restaurant which are inturn provided by the RSU's, they send requests to the nearest RSU. The figure 3 provides a flowchart which presents the process of vehicles requiring services from the nearest RSU\cite{qu2015security}.
\begin{figure}[h!]
	\centering
	\includegraphics[scale=0.70]{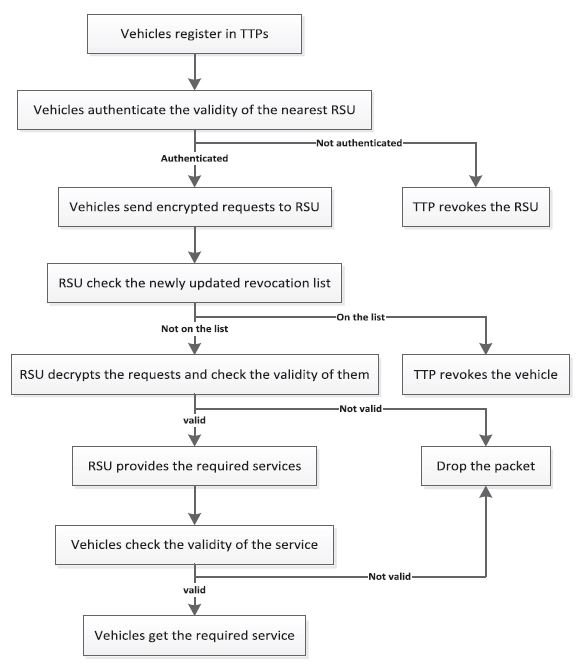}
	\caption{The process of vehicles requiring services from the nearest RSU[11].}
	\label{fig:The process of vehicles requiring services from the nearest RSU.}
\end{figure}

\par In most cases, in order to broadcast messages, the vehicles should first get authenticated with RSU and seek the permission\cite{5982404}. When the vehicles passes the RSU which is in its nearest vicinity, it should also authenticate the validity of the RSU in case it is a fake RSU. As soon as the RSU authentication is completed, the vehicle sends the encrypted request messages and its certificate to the RSU. The RSU decrypts the request and then looks up the newly updated revocation list retrieved from TTP to check whether the vehicle is entitled to obtain the service. If the certificate is on the revocation list, the RSU rejects the request, otherwise the vehicle is authenticated.If the vehicle is authenticated, the RSU sends the response back to the vehicle and provide the service request.

\subsection{Evolution of Security}
\par The evolution of security can be well understood by accessing and classifying the security mechanisms of in-vehicles and V2X systems.The classification is based on risk analysis of security technologies used in vehicles at different layers - software, hardware, cryptography, architecture, network technologies, and protocols. Based on the above mentioned layers, we categorize the security mechanisms into:
\subsubsection{Software implementations}ECUs software implementation may consist of design flaws or bugs through which an intruder can easily evade a security mechanism. For example, classical buffer overflow were used in  Heartbleed attack (a security bug) on openSSL, A FREAK  (man-in-the-middle) attack in openSSL that allowed to use  weak easily crackable RSA keys .
\subsubsection{Configuration} According to OWASP Top 10 list\cite{owasp}, the fifth place being the “Security Misconfiguration” is the most critical web application security flaw. Usually security mechanism are very difficult to configure due to its high complexity. There are high chances of committing mistakes, for example firewall configuration, intrusion detection rules, or access control lists. Some of the common misconfigurations are displaying error handling messages, directory listing in web servers, misconfiguration of firewall rules, or using default passwords.
\subsubsection{Protocols}Secured communication protocols are known as security protocols. They encapsulate the communication between the two agents. Insecure communication protocols are usually vulnerable and are often exposed to various attacks, such as replay, man-in-the-middle attack, and impersonation. In order to overcome these attacks, defining a security protocol has always been a big challenge, which leads to false assumptions resulting in a potential attack. For example, Wired Equivalent Privacy (WEP) protocol, used to protect from eavesdropping and other attacks securing the link-layer communication, has several security flaws in the protocol, deriving from mis-application of cryptographic primitives\cite{Borisov:2001:IMC:381677.381695}.
\subsubsection{System Security}In ECUs, the system security mechanism deployed is inadequate. If more complicated attacks become available, Address Space Layout Randomization may fail. 
\subsubsection{Symmetric cryptography}Attacks against cryptographic ciphers and hash functions may become more powerful or  robust, but as far as the security is concerned it might not be as secure as it was couple of years ago. Furthermore, even if the algorithm may be very secure but the cryptographic key length may be the loophole for vulnerability in the coming years\cite{Stevens2016}. For example, in construction of a quantum computers, most of the implementations use keys of 128-bit, as per the Grover’s algorithm\cite{Grover:1998:FFQ:276698.276712} it is required to change the key length to 256-bits which in turn reduces the bits in half.
\subsubsection{Asymmetric cryptography}The main difference between symmetric and Asymmetric cryptography is the mechanism  and the key lengths (longer). For example, Many servers used a single 512-bit group for Diffie-Hellman key exchange . But later it was improvised by using 2048-bit or longer primes to avoid Logjam Attack on TLS\cite{Adrian:2015:IFS:2810103.2813707}.
\subsubsection{Hardware security modules}In most of the above mentioned mechanism replacing a hardware is not easy due to their hardware being immutable and performance and various cost reasons. Even if hardwares with higher flexibility like FPGAs are used there will be a huge difference in their performance of powerful mechanisms. 

\section{Background}
\par For most of us, we think the beginning of smart cars as a very new concept in the world of automobile. A humble beginning of smart car started  in early 1990’s when the Swiss watch company Swatch, the founder Nicolas Hayek came up with the idea of Smart cars and then ended up partnering with Mercedes Benz and the concept of smart car was born.  The name “smart” comes from the partnership between Swatch and Mercedes (S \& M) and the fact that they wanted to design an “artistic” little car. In 1994, Mercedes-Benz engineers started with a full development concept car and the smart fortwo vehicle was launched in the year 1997 at Frankfurt Motor Show. 

\subsection{Definitions of smart cars}
\par Though there are many different definitions of smart cars available in the literature, yet there is no officially all-accepted definition present till date. As per SAE International Standard J3016\cite{onsae}, which provides the taxonomy and definitions for automated driving in order to facilitate collaboration among technical and policy domains.Figure 4 depicts the levels of automation defined by SAE J3016.

\begin{figure}[h!]
\centering
\includegraphics[scale=3.50]{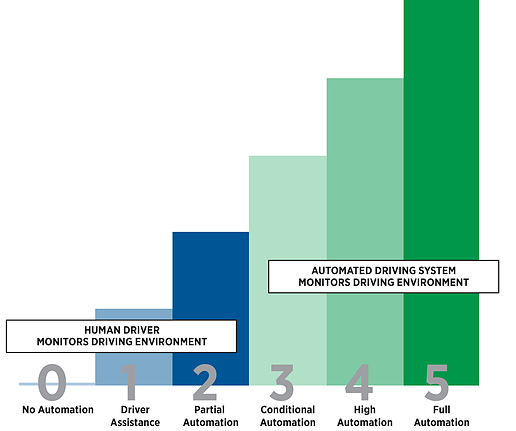}
\caption{Automation Levels of Vehicle\newline \footnotesize source:www.sae.org/autodrive}
\label{fig:Automation Levels of Vehicle}
\end{figure}

\par Another definition comes from the Declaration of Amsterdam under Cooperation in the field of connected and automated driving\cite{regjeringen.no} which lists out major distinction between connected cars involving communication between vehicles and also with the infrastructure (CITS) and automated driving involving use of on-board sensor, cameras, navigation system and various associated softwares.

\subsection{Architecture of a smart car}
\par Defining a generalized architecture for smart cars is not possible. It various from vehicle to vehicle. The architecture of subnetworks and protocols differ from vehicle to vehicle. Hence figure 5 depicts an abstract high-level architecture of a smart car system\cite{enisa.europa.eu}. 

\begin{figure}[h!]
\centering
\includegraphics[scale=0.40]{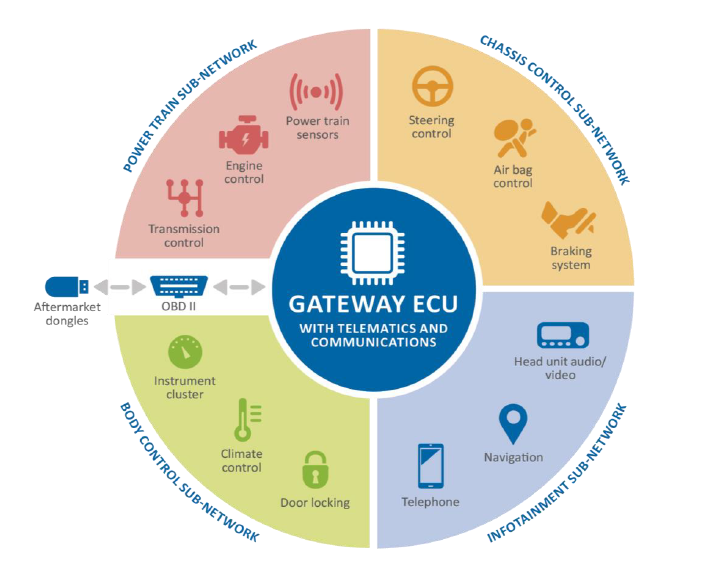}
\caption{ High-Level architecture of a smart car [19]}
\label{fig: High-Level architecture of a smart car}
\end{figure}

\par Majority of the car architectures contains a central gateway Electronic Control Unit (ECU) which are interconnected with different domains. These domains are distinct independent features of the car. These entities are prone to potential risks considering the amount of diverse domains they are interconnected to. The influence of these risks may differ between safety, security or privacy concerns.

\subsection{Assets of a smart car}
\par The car architectures differentiate between various domains (components) which are interconnected with a central gateway (as shown in figure 6).These components may cause risks. This is the reason why components of a smart car are described as assets which needs a stronger protection.

\begin{figure}[h!]
\centering
\includegraphics[scale=0.40]{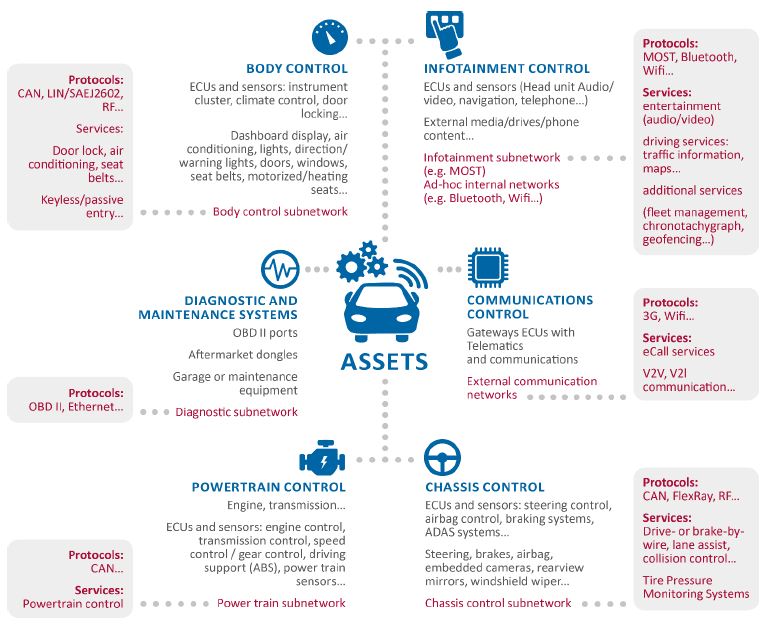}
\caption{ High-Level architecture of a smart car [19]}
\label{fig: High-Level architecture of a smart car}
\end{figure}
\par  Components of a smart car can be categorized into following:
\subsubsection{Powertrain Control}This domain takes care of the link between the car’s energy source to the car until motion.
\subsubsection* {ECUs and sensors} 
The electronic and the mechanical part (powertrain, brake, airbags etc.) of today’s modern cars are controlled by Electronic Control Unit (ECU) . Different domains have different ECUs.Automotive devices usually depends on the ARM platform for application processors, some of the less likely used architectures are SH,V850 and TriCore\cite{ARM}. The automotive-grade processors are more preferred than commercial-grade processors due to the constrained operating environment like temperature, humidity, lifespan in automotive environments. 
\par In particular for vehicular communications, for increasing the level of security, these system must depend on a Trusted Platform Module (TPM) or Hardware Security Module (HSM). ECU applications may depend on real time operating systems like AUTOSAR, Integrity or VxWorks.
\subsubsection* {Subnetwork}
Controller Area Network (CAN) protocol, an ISO standard from 1993 is the widely used and the most popular bus to which most of the ECUs are connected. In a vehicle there may exist many CAN buses to isolate the functions with higher level of criticality like powertrain management from functions with lower level of criticality like multimedia management in a system.The traffic on this bus depends on the solution, at times it can support up to hundreds of messages per second.Hence the CAN bus is the best example and has been analyzed completely by the researchers\cite{miller2013adventures}. The only drawback being some issues with scalability, bandwidth and security.
\subsubsection* {Other components}
This domain deals with the body of the car (physical system) such as combustion engine, transmissions, wheels and steering etc. 

\subsubsection{Chassis Control} This domain takes care of control frame of the vehicle with respect to its environment.

\subsubsection* {ECUs and sensors}
The functionalities of the ECUs are almost the same as that of the powertrain control.They handle the functionalities such as airbag management, steering and brake systems and advanced driver assistance systems. 
\subsubsection* {Subnetwork}
Subnetwork rely on protocols such as FlexRay or RF usually used for Tire Pressure Monitoring Systems. Also it primarily relies on CAN protocol, but the main drawback is, its on a slower side. Flexray or RF is usually designed for drive-by-wire applications which replaces mechanical functions with softwares.
\subsubsection* {Other components}
Other components include embedded cameras, rear view mirrors, windshield wipers and also steering and brake system.

\subsubsection{Body  Control}  This domain takes care of the body of the vehicle, in the sense it takes care of the passenger’s area.

\subsubsection* {ECUs and sensors}
The main functionality on ECUs and sensors is the in-car management system like climate control, auto temperature adjustment, door locking system and instrument cluster. They manage all the required comforts of the passenger. 

\subsubsection* {Subnetwork}
The subnetwork is based on CAN protocol or RF protocols. These RF protocols are used in proprietary protocols like key fobs. They also include other technology like Smart Wave, Bluetooth low energy, Wifi, Zigbee etc. 
\subsubsection* {Other components}
Other components comprises of dash board display, air conditioning, heating seats, warning lights, seat belts, doors, windows, etc.

\subsubsection{Infotainment   Control} This domain takes care of entertainment service installed in the car. This might include navigation system, communication systems (bluetooth telephone access), media system etc.

\subsubsection* {ECUs and sensors}
The core functions include the main unit for audio video content, communication systems, entertainment services, internet connectivity, providing information regarding the current traffic and navigation system through maps. They also manage technical services like fleet management, monitoring of speed and distance (tachograph), creating a virtual geographic boundary using geofencing  (GPS and RFID). 
\par  These additional services makes infotainment ECUs to have a specific architectures. For these systems, even the mobile operating systems can also be used in ECUs (Android, Tizen and WebOS). It is possible to sync the user's smartphone into the vehicle system using QNX. For example, it's used in Apple Carplay and Android Auto technologies which allows the user to view the interface of the phone through the infotainment system installed in the car which in turn helps the user to avoid using his phone while driving. There are many open-source projects that are used to create software solutions for automotive applications. For example, Automotive Grade Linux (AGL) and Linux Genivi.

\subsubsection* {Subnetwork}
The subnetwork depends on ad-hoc networks using Wi-fi or Bluetooth and they mostly rely on wireless connectivity provided either by an embedded UICC or by a smartphone connected by USB cable or with Bluetooth. Additionally, camera system can be connected through an Ethernet.  
\subsubsection* {Other components}
Other components that can be considered as an asset as all the external media devices like drives and phones.

\subsubsection{Communications Control} This domain is a set of communication features offered by a Telematics Control Unit (TCU),which acts a gateway. This domain is completely different of that of the previous ones. 
\subsubsection* {Gateway ECUs}
\par Gateway collects the data from many ECUs using the buses available in the vehicle system and provides Internet connectivity remotely by using driver’s smartphone or through an embedded GSM. Gateways provides all the security protections like authentication and firewall features required for the communication.
\par Some of the use cases that use TCU connectivity are as follows:
\begin{itemize}
\item Remote engine ignition
\item Tracking stolen vehicles
\item eCall- Emergency Warnings (compulsory in Europe from 2018)
\item Remote transmission of vehicle data
\item Smart driving Assistant (e.g.improve driving habits and fuel efficiency)
\item Eco-driving
\end{itemize}

\subsubsection* {External communication networks}
Different kinds of services are provided by TCU which typically involves Wifi and 3G connectivity (V2X communication). The figure 7 provides an overview of external interfaces that a car consists of. These interfaces are usually for long range communication, as well as wired or wireless interfaces.

\begin{figure}[h!]
\centering
\includegraphics[scale=0.40]{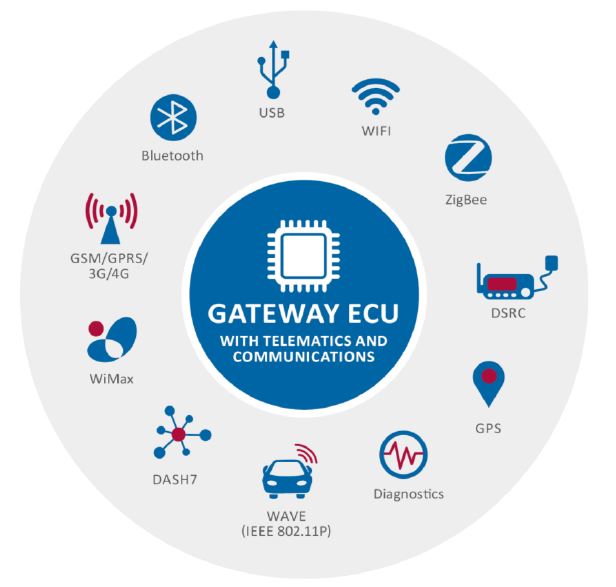}
\caption{ External interface of a smart car [19]}
\label{fig: External interface of a smart car}
\end{figure}

\par TCU Provides various wireless protocols in addition to wired protocols such as USB or any diagnostic protocols. Some of the wireless protocols that TCU provides are as follows:

\subsubsection* {Long-range wireless protocols}
Telematics depends on wireless connectivity such as 2G, 3G and 4G which are catered by the driver’s mobile phone. Some of the mobile protocols such as GSM, GPRS, 3G, 4G, UMTS and LTE are used in different scenarios. GNSS is used as the localization feature in smart cars.

\subsubsection* {Intra-vehicle wireless protocol}
The widely used protocols for intra-vehicular communication are bluetooth and wifi. In recent times there is a transition from the tradition protocols (bluetooth and wifi) to state-of-art protocols such Zigbee, Passive RFID, UWB or 60 GHz mm Wave\cite{6823640}. For example, DASH7 used for Tire Pressure Monitoring System (TPMS) can be used for communication with sensors from near-range protocols to long-range protocols. For example, Mirrorlink, CarPlay or Automotive Link uses wifi or bluetooth to communicate with smartphones using dedicated protocols. Wearable and smart home devices also gets benefited from these type of interfaces ( Open Connectivity foundation project)\cite{openconnectivity.org}.

\subsubsection* {Vehicle -to-Infrastructure (Inter-Vehicle) wireless protocols}
For ITS communications, Inter-Vehicle communications uses Dedicated Short Range Communication (DSRC) that has a bandwidth of 5.9 GHz.  Such communication uses protocols such as WAVE (Wireless Access in Vehicular Environment) which is a means of operation used in IEEE 802.11.Possible alternatives that the state-of-art research provides are DSA\cite{6823640}, WiMAX for V2I communication\cite{5558216} or CEN-DSRC for Electronic Tolling.

\subsubsection{Security, Safety and Privacy concern}
\par Assets are related to safety in several ways:
\begin{itemize}
	\item Compromising powertrain or chassis ECUs and networks may obviously cause a vehicle to behave in an unexpected way, for example if an attacker illegitimately compromises ignition, steering, brakes, speed and gear control, or driving support (such as ABS)\cite{autopilot}
	
	\item Compromising body ECUs and networks systems that may increase harm to the passengers, should they malfunction:
		\begin{itemize}
			\item airbag or safety belts,
			\item door force-lock used for child protection,
			\item the windshield wipers,
			\item air conditioning,
			\item motorized or heating seats,
			\item automatic trunk closing
		\end{itemize}
	
	\item Infotainment ECUs and networks may also cause safety issues : incorrect navigation data may lead the car to unsafe areas, and a disturbance of the audio in the entertainment system (such as high volume burst) may distract the driver.
		
\end{itemize}

\subsection{Overview of Intelligent Transportation System}

\begin{figure}[h!]
\centering
\includegraphics[scale=0.40]{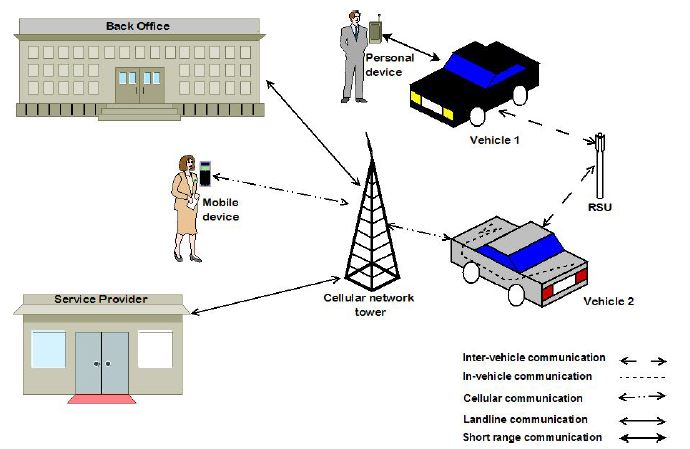}
\caption{ Example of Intelligent Transportation System [2]}
\label{fig: Example of Intelligent Transportation System}
\end{figure}
\par The above mentioned figure comprises of the following entities:

\begin{itemize}
	\item Connected Cars
	\item Back office (e.g. Fleet Management System)
	\item Service Provider (e.g. e-call service)
	\item Personal Device
	\item Mobile Device
	\item Road-Side Units (RSUs)
\end{itemize}

\par Each car exchanges the messages in in-vehicle network through ECUs that regulates the functionalities of the vehicle. For example, vehicle dynamics control system (VDCS) of a vehicle uses the angle of steering wheels and other relevant informations to assist the driver in over-steering, under-steering and roll-over scenarios\cite{alur2007handbook} .
\par Cars exchange messages between each other through VANets. A standard for inter-vehicle communication-Vehicular Environments (WAVE)\cite{4939288} uses wireless access to communicate with the nodes of a VANet technology. The exchange of information plays such a vital role in order to avoid collision. With the help of WAVE, vehicles communicate with their neighbouring RSUs in order to provide information about its own activities like current location, speed at which the vehicle is moving and the direction of the vehicle and get the feedback of the ongoing traffic updates based on the location. 
\par Vehicle 1 interacts through WPAN with a personal device and driver can remotely start his vehicle by using his PDA. With the means of cellular network, vehicle 2 interacts with Back Office (Fleet Management System) by sending its data about its location and communicates with Service provider (eCall) in case of some emergency\cite{eCall}.
\par This paper provides an overview of identifying different types of security and privacy attacks, and classify their defending mechanisms in VANETs. Section III provides an overview of different security attacks in VANETs, their security attributes and types of different malicious vehicles. Further, section IV includes various security attacks and their defensive approaches. Section V provides security implications that a vehicular network possess and Section VI provides a discussion and summerization of the entire paper with respect to security attacks and privacy ussues in vehicular network.

 \section{Security in Vehicular ad hoc networks}
 \par Vehicular Ad hoc Networks are the latest mobile ad hoc network which consists of mobile routing protocols for communication between vehicle for an intelligent transportation systems. Considering the frequent movements of the vehicle and possessing a hybrid architecture, their security and privacy has been a major concern among researchers. Hence designing a secure platform to authenticate and validate the transmitted messages among the vehicles is very important in VANETs. 
\par In recent times, VANETs has been very popular and widely used in many application in automobile industry such as traffic control and monitoring, toll system, highway internet access, improving safety of the highway system. VANETS are also popularly knows as Wireless Access in Vehicular Environment (WAVE)\cite{4526014}. WAVE uses Dedicated Short-Range Communication (DSRC) in Intelligent Transportation System (ITS)\cite{1580935}.
\par The security in VANETs plays a very vital role as it deals with the life critical application and should attain the security requirements such as privacy, Integrity and confidentiality to make the system free from vulnerability and secure against malicious attackers.Security attacks such as Denial of Service (DOS)\cite{4365481}, Wormhole attack\cite{5403263}, Sybil attack\cite{4428742} and Purposeful attac \cite{5189734} attacks drivers privacy which leads to loss of life. The main goal of today’s research is to provide a secure communication channel where the identity of the driver is not exposed, if failed, the attacker may consume the data for setting up the attacks with fake identities without even getting caught. But the main challenge of drivers and vehicle is disclosing the identity to RSUs in order to establish the communication. Hence obtaining a secure channel is very important and security and privacy should be handled with at most care.
\subsection{Security Attributes}
\par  There are many requirements to achieve security in VANETs, which are discussed as follows\cite{Raya:2005:SVA:1102219.1102223}.
\subsubsection{Authentication} It is very important for a vehicle to respond to a message sent an authenticated source. Vehicles should always respond to the messages sent by a genuine member of the network.
\subsubsection{Data Verification} As soon as the authentication is done for the sender vehicle, the receiving vehicle verifies the data to check whether the messages received contains genuine or corrupted data. 

\subsubsection{Availability}The performance should not be affected when the network is under an attack. Hence, even if the network is under attack it should be available by using an alternative mechanism. 

\subsubsection{Data Integrity} It makes sure that the data is not altered by the attackers. Best example for data integrity is Man in the middle attack, i.e. (In figure 9) If a vehicle B sends a “Road Clear” message to a malicious vehicle C and C changes the message as “Traffic Jam Ahead” and sends the message to vehicle D, thus causing problems to the legitimate vehicle D.

\begin{figure}[h!]
\centering
\includegraphics[scale=0.80]{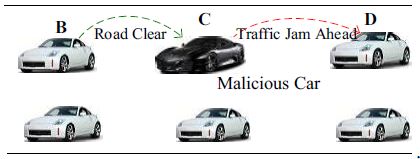}
\caption{  Data Integrity [39]}
\label{fig:  Data Integrity}
\end{figure}

\subsubsection{Non-repudiation} Whenever an identity or an investigation of a vehicle is required, sender should not deny the message transmission at any cost. 

\subsubsection{Privacy} The driver’s personal information should be kept out of reach against unauthorized access. 

\subsubsection{Real-time constraints} Should always maintain real-time constraints as vehicles are connected to VANETs for a short duration of time.

\subsection{Types of Attacker's Vehicle}
\par In VANETs, malicious vehicles attacks legitimate vehicles in different ways. Capacities on an attacker can be classified as below: 

\subsubsection{Insiders Vs Outsiders} Member nodes who communicate with each other inside the network are called as Insider. Insider is an authenticated member of the network, always possess a certified public key and can have different varieties of attacks. Outsiders are the external entity who cannot communicate directly with the members of a network i.e. they are considered as an intruder by the members of the network and have less variety of attacks.

\subsubsection{Malicious Vs Rational} A malicious attackers are less predictive and attacks a network without any personal benefit. In the process it damages various member nodes and the network. On the other hand, rational attackers are highly predictive and always follow a pattern. They always perform attacks for some personal benefits.

\subsubsection{Active Vs Passive} An active attacker can generate new packets to damage the network where as a passive attacker can only listen to the wireless channel but cannot generate the packets. Passive attackers are relatively less harmful than the active attacker.


\section{Security Attacks and defensive mechanism}
\par This  section provides an overview of various security attacks that takes place on Vehicular Ad hoc Networks (VANETs) and proposes the defencive mechanism to overcome these attacks\cite{Douceur:2002:SA:646334.687813}\cite{4428742}\cite{raya2005hubaux} \cite{4437823}\cite{Newsome:2004:SAS:984622.984660}\cite{6079000}

\subsection{Bogus Information} 
\par  An attacker can inject wrong (bogus) information into the network of their choose. These attacks are generally related to authentication security requirement.  For example, an attacker can feed wrong information about the traffic conditions in the network in order to make its movement fast and easier on the road. 
\par A message authentication scheme called as Elliptic Curve Digital Signature Algorithm (ECDSA)\cite{5189734} keeps the message secure and provides a strong authentication for the destination vehicles by a technique called hashing. This message authentication scheme works by generating public and private keys from the source vehicle. Public key is available in all the vehicles present in the network. With the help of hash algorithm and private key, the source vehicles hashes the message and send the encrypted message to the destination vehicle. The destination vehicle then decrypts the message using the public key. Changes made in messages will also change in the hash messages making this scheme a unique one.  Hashing being a strong technique makes this scheme more secure on message authentication.

\subsection{Denial of Service (DoS)}
\par Efficiency and performance of a network can deteriorate when the attackers transmit dummy messages to jam the channel. Figure 10 illustrates the operation on DoS where a malicious black car creates a dummy message “Lane close Ahead” to RSU and another vehicle behind it to create a jam in the network. Figure 11 demonstrates a Distributed DoS where the attack takes in a distributed manner. Many malicious cars attacks the legitimate cars in the network. V1 is attacked by many malicious cars from different locations at different time.Hence V1 cannot establish a communication to other vehicles in the network.

\begin{figure}[h!]
\centering
\includegraphics[scale=0.65]{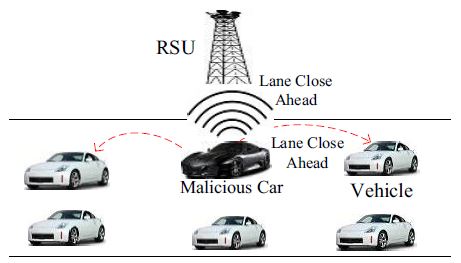}
\caption{ Denial of Service (Dos) Attack [39]}
\label{fig: Denial of Service (DoS) Attack}
\end{figure}

\begin{figure}[h!]
\centering
\includegraphics[scale=0.65]{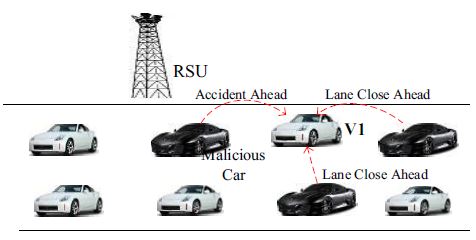}
\caption{ Distributed Denial of Service (Dos) Attack [39]}
\label{fig: Distributed Denial of Service (Dos) Attack}
\end{figure}

\subsection{Masquerade}
\par With the help of message fabrication, alteration and replay a vehicle fakes its identity to be another vehicle for its personal advantage. For example, a malicious vehicle can fake around as an ambulance vehicle to defraud other vehicle to slow down the ongoing traffic. 

\subsection{Black Hole Attack}
\par Black hole attack is an attack where many malicious cars acts as a barrier between the other (legitimate) cars by changing the path of the messages in a network. A black hole is a part of network where the network traffic is redirected. The black hole area usually does not consist of any nodes, even if it exists they are less likely to be participated in the network resulting in the loss of data packets.

\begin{figure}[h!]
\centering
\includegraphics[scale=0.75]{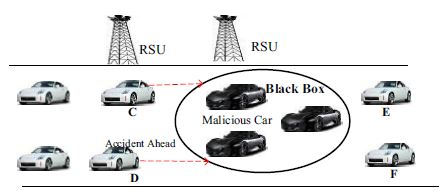}
\caption{ Black hole Attack [39]}
\label{fig: Black hole Attack}
\end{figure}

\subsection{Malware and Spam}
\par Viruses and spams can cause a huge negative impact on a VANETs operations. These attacks are usually initiated by insiders rather than outsiders whenever there is an update performed by on board units (OBU) and road side units (RSUs). These attacks causes an increase in the latency of a transmission which can be made less severe by using a centralized administration. 

\subsection{Timing Attack}
\par Data Integrity and security is achieved by transmitting the data from one vehicle to the other at the right time. In this attack when a malicious car receives any emergency message they do not forward it to the neighbouring vehicle at the right time, instead they add some timeslots to the message to create some delay in transmitting it. Hence neighbouring vehicles receive message when it's too late to act upon that emergency situations.

\begin{figure}[h!]
\centering
\includegraphics[scale=0.70]{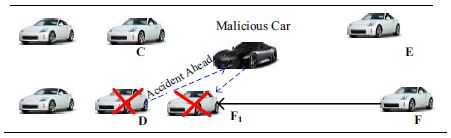}
\caption{ Timing Attack [39]}
\label{fig: Timing Attack}
\end{figure}

\par The above figure depicts that when the car D sends “Accident Ahead” message to the malicious black car, it does not forward the message to its neighbouring vehicle at the right position F but transmits by adding some timeslots so that when the vehicle receives its message its position would be at F1 where the emergency situation has already taken place. 

\subsection{Global Positioning System (GPS) Spoofing}
\par A location table that contains a log of geographic locations and vehicles identity in the network is maintained by a GPS satellite. An intruder can inject false reading in GPS positioning system to mislead vehicles to think that they are in a different location. These attackers uses GPS satellite simulators as a medium to produce stronger signals than those of the actual ones. 

\subsection{Man in the Middle Attack (MiMA)}
\par A malicious vehicle listens to the other communication in the network and inject false information between the vehicles. Figure 14 illustrates MiMA attack where the malicious black car interferes in between B and D and also, sends the false information to the car E, (original message) that was received from the car A.

\begin{figure}[h!]
\centering
\includegraphics[scale=0.75]{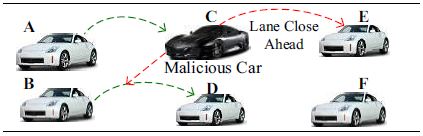}
\caption{ Man-in-the-Middle Attack [39]}
\label{fig: Man-in-the-Middle Attack}
\end{figure}

\subsection{Sybil Attack}
\par In this attack each node send messages with several identities. The attacker creates several identities to simulate several nodes. Hence other vehicles in the network assumes that there are several vehicles in the network at the same tim \cite{Boneh:2001:IEW:646766.704155}\cite{5379844}. This problems leads to a catastrophe where a vehicle can claim to be at many positions at the same time creating a huge security risk in the network. 
\par In Sybil attacks, to resist these threats many systems uses the concept of redundancy by limiting the physical entities to some resource\cite{Douceur:2002:SA:646334.687813}\cite{Newsome:2004:SAS:984622.984660}. Sybil attacks can be detected through resource testing. The research contribution done in\cite{Douceur:2002:SA:646334.687813} to test the computational resources of each node is done through computational PUZZLES. But, this approach is not feasible for VANETs\cite{Newsome:2004:SAS:984622.984660} because an attacker node can consist of more computational resources compared to an ordinary node. This problem can be avoided by radio resource testing\cite{Newsome:2004:SAS:984622.984660}.
\par Another mitigation strategy is using public key cryptography where in each vehicle is authenticated using public keys\cite{4015703}. Predefined propagation model is another approach used in detecting Sybil attacks in wireless networks\cite{Xiao:2006:DLS:1160972.1160974}. In this model the differences of signal strength between sent and the received signals is matched with the claimed position through Received Signal Strength Indicators (RSSI) approach.
\par RobSAD is another approach that detects the Sybil attack based on the normal and abnormal motion trajectories of vehicles\cite{5158865}. This approach detects the attack during the initial deployment phase of VANETs.With the help of RSUs each node can detect attacks on their own. These RSUs can provide vehicles digital signatures with time-stamp regularly on a consistent basis. Each node in the network can keep a log of these digital signatures and compare the differences from their neighbouring nodes signature vectors to detect Sybil nodes. Thus each node can detect attacks own their own by comparing the digital signatures without collaborating with their neighbour nodes resulting in low system requirements and high detection rate. 

\begin{figure}[h!]
\centering
\includegraphics[scale=0.75]{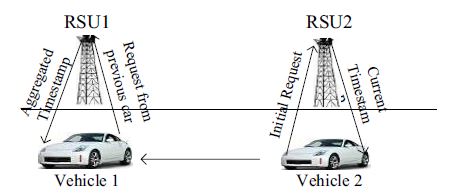}
\caption{ Timestamp series approach [39]}
\label{fig: Timestamp series approach}
\end{figure}

\par Another approach is called the Timestamp series approach and its best to use for an initial development stage of VANETs with an RSU infrastructure. Figure 15 demonstrates the working of timestamp series approach. The RSU provides digital certificates to all the vehicles that comes across it and believes that no two vehicles can pass more than one RSUs at a time. Thus, it is easy to detect if a vehicle receives more than one message with the similar timestamp certificates.

\subsection{Wormhole Attack}
\par Wormhole is a grievous attack in VANETS and other ad hoc networks. A tunnel is created from two or more malicious nodes that acts as a channel to transmit the data packets from one end of the network to the malicious node at the other end. These data packets are broadcasted to the entire network. These short network connections or the links intimidate the security of transmitting data packets and delete them.
\par In VANETs and in on -demand routing protocols (such as AODV or DSR), wormhole attack interrupts the multicast and broadcast messages. In AODV protocol the main loop hole for the attack is not using any authentication and protection mechanisms for routing packets.  Denial of Service (DoS) attacks are performed by unauthorized access gained by wormholes ar malicious node. 

\begin{figure}[h!]
	\centering
	\includegraphics[scale=0.70]{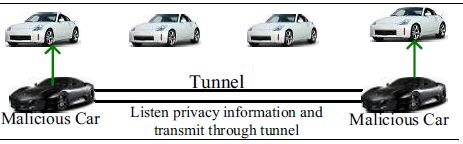}
	\caption{ Wormhole Attack [39]}
	\label{fig: Wormhole Attack}
	
\end{figure}
\par Figure 16 depicts a wormhole attack where the confidential information is transmitted from one end of the network to the other end by the malicious black car.
\par The TIK protocol is a packet leashes based protocol for detecting and defending against wormhole protocol. Packet leash is an approach to overcome wormhole attacks\cite{hu2003packet}. All the nodes are synchronized and with clocks in the network and all the nodes are aware of the clock differences between any two nodes.
\par Asymmetric cryptography is used by TIK protocols to cater immediate authentication of the received packets where it uses \textit{k} public keys for \textit{k} nodes and hash functions for keeping up-to-date keys data and packets that were received. The difference between the packets travel distance and allowed distance to travel allows to detect the presence of an attacker.
\par An improvement of packet leashes method called HEAP was  introduced\cite{5403263} and it is an efficient approach used to detect the wormhole attacks in AODV routing protocol of VANETs. HEAP provides more security and low overhead. To detect malicious node, HEAP uses geographical lashes rather than local lashes. The main difference in geographical lashes is the limitational of packets travel distance . Hence to overcome this problem, whenever their travel distance is more than the value claimed, the packets are automatically dropped.

\subsection{Illusion Attack}
\par As the name suggests an opponent produces an illusion to other vehicles in the neighbour vicinity by broadcasting the traffic warning messages on current road conditions. Illusion Attack is the youngest security threat and not many efficient solutions have been put forward for this attack. Due to the created illusions, the drivers’ behaviour is altered and force them to react to those responses. This may lead to traffic jams, car accidents and also decrease the performance of VANETs. Current authentication methods are not effective against Illusion attack because the intruder manipulates the behavior of the sensors to produce produce false information. 
\par One solution against illusion attack is the Plausibility Validation Network (PVN)\cite{4437823}, which is a new security model for the security of VANETs against illusion attacks. Raw sensors’ data are collected by PVN and verifies whether the data collected are plausible or not. Inputs like incoming data from antennas and sensor datas are classified by an input data header. Validity of the input data is monitored by a rule database and data checking module and takes actions if necessary. If the data passes the validity test, the data is assumed to be a clean one, if not it is considered as invalid.

\subsection{Intentional  Attack}
\par Prevention of Intentional Attack is very difficult as they are authenticated entities that perform peer communications with neighbours. They are trusted entities in the network that communicates with its neighbour in the network.
\par It is very important to protect against the misbehaviour of intentional attackers or unintentional malfunctioning  hardware systems. There are high chances of misbehaving nodes discarding the messages that is received from other nodes, misinterpreting the messages and improper use of bandwidth or injection of bogus messages. A technique is proposed where anonymous communication to protect against misbehavior and promises to keep the privacy of vehicles. This technique is used in vehicle-to-vehicle communication and vehicle-to-infrastructure communication systems\cite{SUN20091515}. A threshold authentication technique is used to authenticate the misbehaviour of malicious nodes by providing a threshold value. Any authentication above the threshold value guarantees in identification of misbehavior node’s details. 

\subsection{Impersonation Attack}
\par In a network  a malicious vehicle delivers the message on behalf of other vehicle to create catastrophe in the network that leads to traffic jams and accidents. The malicious vehicle after sending the message it gets camouflaged and hides itself from the other vehicles in the network. But, usually in vehicle-to-vehicle communications, one vehicle sends the security message to all the other vehicles that may have an impact on the other vehicles and traffic control system. Hence to reduce the communication overhead, all the messages should be signed and authenticated.
\par A research contribution has been published in detection of impersonation attack, called as SPECS\cite{CHIM2011189}. It is a scheme used widely in V2V communication systems which provides secure and privacy enhancing communication schemes for VANETs. This scheme is built on the basis of IBV protocol\cite{4509653} which cannot fulfill privacy requirements due to its suffering from impersonation attack. The security is works as follows:
\par Pseudo-identity and shared secret key $m_{i} $ among the vehicles and RSU is used to safeguard the identity of the vehicles. The scheme uses PKI with its real identity RIDi and password PWDi with nearby RSU to authenticate the vehicle. A shared secret key $m_{i} $ for vehicle and RSU is generated to authenticate the vehicle. This key is generated by RA. TA forwards $m_{i} $ with a hash function and an encrypted block that has $m_{i} $ and system secret key, \textit{s}. Only the authorized user can decrypt the encrypted block.  \par This block is transmitted to the vehicles by RSU. A new shared key is generated whenever a vehicle passes through a new RSU. A shared key and one way hash function with the signing key is used to generate the signature. RSU and TA attackers cannot generate valid signing key to sign the message as $m_{i} $ are known only to a vehicle. Using a batch verification process performed by RSU, it is easy to detect the invalid signatures and attackers. The whole batch is dropped in IBV if any invalid signature is identified by using batch verification process. Whereas in SPEC the whole batch is not dropped, instead it uses binary search where in it divides the batch in two halves and checks the invalidity on each half. When an intruder is found, it informs the other vehicles and continues this process until all signatures are validates. RSU sends the message to all the vehicles excluding the hash value which is inturn stored into positive and negative bloom filters. Received message’s identity is known by creating the hash value and comparing it with the bloom filters has value. If the hash value of a message is found in positive bloom filters, then the message is considered to be valid. Else, the message is discarded and labeled as invalid message.

\subsection{Eavesdropping}
\par This attack happens when an attacker is located in a vehicle, be it stopped or moving, or in a false RSU.The collection of vehicle-specific information from overheard vehicular communications is easy in a wireless network. The attackers obtain the target vehicles’ confidential data, including the drivers real identities, their preferences or even their credit card codes, which seriously violates the privacy of the drivers.

\subsection{Message suspension}
\par This attack happens when adversaries hold onto messages before sending them. An attacker selectively drop packets of messages from the network, which may hold critical information for the intended receiver, and the attacker suppresses these packets and can use them again in the future. One goal of such an attack would be to prevent registration and insurance authorities from learning about collisions involving the attacker’s vehicle and/or to avoid delivering collision reports to roadside access points.

\setlength{\extrarowheight}{5.0pt}
\begin{table} [h]
	\caption{Summary of attacks with their types and their security requirements}
	\label{my-label}
	\centering
	\begin{tabular}{|*{12}{p{2.3cm}|}}
		\hline
		\textbf{Name of Attack} & \textbf{Type of Attack} &\textbf{Security Requirement} \\ \hline
		Bogus Information & Insider & Data Integrity, Authentication\\ \hline
		Denial of Service (DoS) & Malicious, insider, network attack & Availability\\ \hline
		Masquerading & Active, insider & Authentication\\ \hline
		Black hole & Passive, outsider & Availability\\ \hline
		Malware & Malicious, insider & Availability\\ \hline
		Spamming & Malicious, insider & Availability\\ \hline
		Timing attack & Malicious, insider & Data integrity\\ \hline
		GPS Spoofing & Outsider & Authentication\\ \hline
		Man-in-the-Middle & Insider, monitoring attack & Data Integrity, Confidentiality\\ \hline
		Sybil & Insider, network attack & Authentication\\ \hline
		Wormhole/Tunneling & Outsider, malicious, monitoring attack & Authentication, Confidentiality\\ \hline
		Illusion Attack & Insider, malicious & Authentication\\ \hline
		Purposeful attack & Active, insiders, malfunctioning hardware & Authentication\\ \hline
		Impersonation & Insider, network attack & Authentication\\ \hline
	\end{tabular}
\end{table}


\section{PRIVACY IMPLICATIONS}

\par In both wired and wireless networks, privacy has always been a key concern, and many researchers have dedicated their work on tackling this problem. Even so, while the level of privacy could be enhanced, the most ideal situation where the users’ information could never be traced, may never come to fruition. Given the large scale and frequent usage of the internet and cellular networks, small little flaws in the aspect of privacy seem to be acceptable. Still, privacy is a decisive factor in the public’s acceptance of and the commercial deployment of vanets\cite{goudarzi2013systematic}. Leaking drivers’ private profiles could lead to serious consequences. For example, location tracking of any vehicle provides access to past and current locations of the vehicle\cite{sampigethaya2007amoeba}. Once the location history has been accumulated, adversaries could infer the driver’s personal interests and daily routine by combining these data with additional information. The information could then be misused for crimes, such as abductions or automobile thefts.
\par Security has been one of the most challenging problems in VANETs and should be considered along with privacy. To secure the communication in VANETs, the data must be authenticated. Through authentication, the network can be aware of the precise location of a specific user at a specific time, which ensures that the TTP could intervene in the vehicle when an issue arises. For example, when a vehicle has an accident on the road and leaves the scene, the TTP could reveal the real identity of the vehicle and track it until the police were able to catch up with the responsible vehicle. However, some drivers are not willing to let the TTP have access to their confidential information. Therefore, how to preserve privacy while still enabling authentication has become one of the main challenges
of implementing VANET\cite{xi2007enforcing}.

\par The automobile has gradually evolved from an analog machine with mechanical components to an electronic system with a growing number of computer-based systems. Within this revolution of “smart car”, GPS vehicle navigation has been the main focus and attention. There are efforts underway to use GPS vehicle navigation infrastructure for additional value-added services like mobility pricing of insurance\cite{norwichunion.com}, infrastructure-less electronic toll collection, and GPS enabled parking fee collection\cite{Bgrush}.

\par These applications would require disclosure of positional data by its users in real time. Systems would process the positional data to charge the motorist for the services rendered. A decrease in the cost of electronic storage means that this captured data intended for a specific purpose, transaction processing, may be retained indefinitely or at least for long periods of time. Since GPS data is information rich, the temptation to use it for secondary purposes may be too great to resist.

\par While theoretical research has tried to raise awareness about these threats and has proposed algorithms to protect the privacy of individuals\cite{Duckham2005}, limited research has been conducted to assess these threats in a real-life scenario.

\section{Conclusion}

\par This paper provides an overview of various vehicular network architectures. The evolution of security in modern vehicles. Various security and privacy attacks in VANETs with their defending mechanisms with examples and classify these mechanisms. It also provides an overview of various privacy implication that a vehicular network possess. Finally, this paper concludes with discussions and summarization of all the security attacks and the privacy issues that a vehicular network consists.

\bibliography{main}
\bibliographystyle{ieeetr}

\end{document}